\documentstyle[12pt]{article}
\begin{document}
\def\simgeq{\; \raisebox{-0.4ex}{\tiny$\stackrel
{{\textstyle>}}{\sim}$}\;}
\def\simleq{\; \raisebox{-0.4ex}{\tiny$\stackrel
{{\textstyle<}}{\sim}$}\;}
\def\si{\sigma _I}
\def\sm{\sigma _M}
\def\sc{\sigma _C}
\font\tencaps=cmcsc10 \def\caps{\tencaps}
\baselineskip=24 pt
\centerline{{\large Analytic Scaling Functions Applicable 
to Dispersion Measurements}}
\medskip
\centerline{{\large in Percolative Metal--Insulator Systems}}
\medskip
\centerline{{\caps D.S. McLachlan, W.D. Heiss, C. Chiteme and Junjie Wu}}
\medskip
\centerline{{\sl Department of Physics, University of the Witwatersrand}}
\centerline{{\sl PO Wits 2050, Johannesburg, South Africa}}

\vskip 1cm


\begin{abstract}

Scaling functions, 
$F_+(\omega/\omega _c^+)$ and $F_-(\omega/\omega _c^-)$ for
$\phi >\phi _c$ and $\phi <\phi _c$, respectively, are derived from an 
equation for the complex conductivity of binary conductor--insulator 
composites.  It is shown 
that the real and imaginary parts of $F_{\pm}$ display most
properties required for the percolation scaling functions. One difference is
that, for $\omega /\omega _c<1$,
$\Re F_-(\omega/\omega _c)$ has an $\omega $-dependence of $(1+t)/t $
and not $\omega ^2$ as previously predicted, but never conclusively observed.
Experimental results on a Graphite-Boron Nitride 
system are given which are in reasonable agreement with the 
$\omega ^{(1+t)/t}$ behaviour for $\Re F_-$. Anomalies in the real dielectric
constant just above $\phi _c$ are also discussed.

\end{abstract}

\section{Introduction}

The ac and dc conductivities of resistor and resistor-capacitor (R-C) 
networks and continuum conductor-insulator composites have been extensively 
studied for many years. In systems where there is a very sharp change 
(metal-insulator transition or MIT) in the dc conductivity at a critical 
volume fraction or percolation threshold denoted by $\phi _c$,
the most successful model, for
both the dc and ac properties, has proved to be percolation theory.   Early 
work concentrated on the dc properties, but since it was realized  
\cite{1.}--\cite{6.} that 
the percolation threshold is a critical point, and that the percolation 
equations could be arrived at from a scaling relation, several papers, which 
are referenced and discussed in a previous paper \cite{7.}, reporting 
experimental results on the ac conductivity have appeared. Review articles, 
containing the theory and some experimental results, on the  complex ac 
conductivity and other properties of binary metal-insulator systems include 
\cite{8.}--\cite{10.}.

In another paper, extensive  dc conductivity and low frequency dielectric 
constant results on systems based on Graphite (G) and hexagonal Boron 
Nitride (BN), which have what are probably the cleanest and sharpest dc MITs 
yet observed in a continuum system,
have been reported \cite{11.,12.}. These G-BN results were
found to obey the percolation equations, as a function of volume fraction 
in \cite{11.,12.}.  In \cite{7.} it was shown that the experimental
dispersion results, for samples with
various volume fractions of G, can be scaled onto two curves which are
consistent with previously measured percolation parameters \cite{11.,12.};
one curve refers to the real conductivity
above and the other to the imaginary conductivity (real dielectric
constant) below the critical volume. Unfortunately, the 
parameters $\omega _c^{\pm}$ that had to be used to achieve this scaling were 
found to be different from those expected from the scaling models (R-C 
lattice \cite{8.} and anomalous diffusion \cite{13.}), no matter
whether the $\omega _c^+$ and $\omega _c^-$
were calculated using the accepted universal parameters or 
the already reported dc conductivity and low frequency dielectric results 
\cite{11.,12.}.  However, there is agreement between the critical exponents, 
characterising the frequency dependence of the real conductivity when 
$\phi >\phi _c$ (where $\phi $ denotes the conducting volume fraction),
the imaginary conductivity for $\phi <\phi _c$ and
the exponents found from the previously reported dc and low frequency ac
results \cite{11.,12.}.

One measurable scaling function that did not agree with the previous 
power law predictions \cite{8.}--\cite{10.} was that for the loss component
in the insulating region. As these results were somewhat controversial they
were not discussed in \cite{7.}. In the meantime further measurements have 
been made on one of the $G-BN$ systems \cite{11.,12.} and other systems, 
using a recently acquired dielectric spectrometer system
which is able to measure far smaller dielectric loss parameters as well as
loss and phase angles. Some of these results are presented in this paper.

The present paper introduces scaling functions which can depend on
{\it complex} conductivities. They closely fit results of the
medium conductance $\sm $, in particular the frequency ($\omega $)
dependence of the first order real part for $\phi > \phi _c$ and the first
order imaginary part for $\phi < \phi _c$ as was shown
in \cite{7.}. Most of the scaling power laws given in \cite{8.}-\cite{10.} 
are obtained, the range of parameters over which these functions can be 
expected to generate accurate scaling functions is derived. It is also shown 
that, while the second order terms of
the scaling functions have the exponents $t/(s+t)$ for $\phi \approx
\phi _c$ and $\omega /\omega _c\gg 1$ as given in \cite{8.}-\cite{10.}, the
second order exponents for low frequencies and $\phi \simgeq 0$ or
$\phi \simleq 1$ differ from those of \cite{8.}-\cite{10.}. New experimental
results for $\sm (\omega ,\phi )$ are presented which agree reasonably 
well with the exponents predicted by the new scaling functions,
provided that the {\it complex} conductivities of the dielectric
components of the continuum systems are taken into account. The measurements
for $\sm $ ($\phi <\phi _c$) in the G-BN systems are definitely not in accord
with the $\omega ^2$ prediction given in \cite{8.}-\cite{10.}.

A feature of the new experimental results is that, where
measurable, the real dielectric constant continues to increase with
$\phi $ for $\phi >\phi _c$ and certainly does not decrease according to
$(\phi -\phi _c)^{-s}$ as given in \cite{8.}-\cite{10.}. However this
increase is qualitatively consistent with the expressions introduced in this 
paper. The effect is more clearly observed in carbon black--polyethylene 
compounds \cite{14.} and in 3D-systems where various conducting powders are
distributed on the surfaces of large insulating grains \cite{15.}.

\section{Theory}

The equation 
\begin{equation}
{(1-\phi )(\si ^{1/s}-\sm ^{1/s})\over
(\si ^{1/s}+A\sm ^{1/s})}+
{\phi (\sc ^{1/t}-\sm ^{1/t})\over
(\sc ^{1/t}+A\sm ^{1/t})}=0
\label{gem}
\end{equation}
gives a {\it phenomenological} relationship  between $\sc $, $\si $ and
$\sm $ which are the conductivities of the conducting and 
insulating component and the mixture of the two components, 
respectively \cite{7.,16.}. Results obtained from an earlier version
\cite{17.} of Eq.({\ref{gem}) are reviewed in \cite{18.}
and references therein, where the dc conductivities of some two phase systems
are successfully modelled  for $s=t$. The conducting volume fraction
$\phi $ ranges between 0 and 1 with $\phi =0$ characterising
the pure insulator substance ($\sm \equiv \si $) and $\phi=1$ the pure
conductor substance ($\sm \equiv \sc $). 
The critical volume fraction, or percolation threshold, is denoted by 
$\phi _c$, where a transition from an essentially
insulating to an essentially conducting medium takes place. 
We use the notation $ A=(1-\phi _c)/ \phi _c. $
For $s=t=1$ the equation is
equivalent to the Bruggeman symmetric media equation \cite{16.,17.}. The 
equation yields the two limits
\begin{equation}
\label{limm}
|\sc |\to \infty :\qquad
\sm =\si {\phi _c^s\over (\phi _c-\phi)^s} \qquad \phi<\phi _c\\
\end{equation}
\begin{equation}
\label{limp}
|\si |\to  0:\qquad
\sm = \sc {(\phi -\phi _c)^t\over (1-\phi _c)^t} \qquad \phi >\phi _c
\end{equation}
which characterise the exponents $s$ and $t$. Note that Eqs.(\ref{limm})
and (\ref{limp}) are the normalised percolation equations.
For ac measurements, \cite{8.,9.,10.} have given equations for the case 
where $\sc $ is real and $\si =-i\omega \epsilon_0
\epsilon _I$, which characterizes a lossless dielectric. However, we
note that all three quantities $\si ,\,\sc $ and $\sm $ can in principle
be complex numbers in Eq.(\ref{gem}). A solution
for $\sm $ can be obtained after rewriting Eq.(\ref{gem}) using the
variable $z=\sm ^{1/t}$, {\it viz.}
\begin{equation}
Az^{1+\alpha }-z^{\alpha }(A\phi+\phi -1)\sc ^{1/t} -z (A-A\phi -\phi )
\si ^{\alpha/t}-(\si ^{\alpha }\sc )^{1/t}=0
\label{expl}
\end{equation}
with $\alpha =t/s$.
We note in passing that Eq.(\ref{expl}) has explicit solutions for
$\alpha =1,2$ and 3, while numerical solutions are easily obtained for larger
integer values. Our interest is now focussed on the question as to what
extent the solution for $\sm $ can be used to obtain valid scaling functions.

The scaling conditions, which are based on those given in
\cite{8.,9.,10.}, read
\begin{eqnarray}  
\sm &=&\sc {(\phi _c -\phi)^t\over \phi _c^t}F_-(x_-),
\qquad \phi <\phi _c   \label{smless}  \\
\sm &=&\sc {(\phi -\phi _c)^t\over (1-\phi _c)^t}F_+(x_+),
\qquad \phi >\phi _c    \label{smplus}
\end{eqnarray}
where the scaling functions $F_{\pm }(x_{\pm})$ depend on the scaling
parameters
\begin{equation}
x_-={\si \over \sc }{\phi _c^{s+t}\over (\phi _c -\phi )^{s+t} }
= -i{\omega \over \omega _c^-}, \quad
\phi <\phi _c  
\label{xsm} \end{equation}
with
$$ \omega _c^-={\sc \over \epsilon_0\epsilon_I}
{(\phi _c-\phi)^{s+t}\over \phi _c^{s+t}}, $$
and
\begin{equation}
x_+={\si \over \sc }{(1-\phi _c)^{s+t}\over (\phi -\phi _c)^{s+t} }
= -i{\omega \over \omega _c^+}, \quad
\phi >\phi _c.
\label{xsp}
\end{equation}
with
$$ \omega _c^+={\sc \over \epsilon_0\epsilon_I}
{(\phi -\phi _c)^{s+t}\over (1-\phi _c)^{s+t}}.  $$
The expressions involving $\omega _{\pm }$ assume specifically a purely
real $\sc $ and imaginary $\si $.
To ensure that curves drawn for $F_{\pm}$ fall on
top of each other for different $\phi _c$,
the normalisation employed in all the equations used in this paper
differs somewhat from the one used in \cite{8.,9.,10.}.
Using the variable $u=F_-^{1/t}$ an equation is found for $u$ by the
substitution $z=u\sc ^{1/t}(\phi _c -\phi)/\phi _c$ in Eq.(\ref{expl}).
It reads for $\phi <\phi _c$
\begin{equation}
Au^{1+\alpha}+u^{\alpha }-
u{(\phi_c-\phi )(1-\phi -\phi _c)\over \phi _c^2}x_-^{1/s}-
x_-^{1/s}=0.
\label{xm}
\end{equation}
In a similar way, the substitution $z=u\sc ^{1/t}(\phi -\phi _c)/(1-\phi _c)$
leads to an equation for $F_+^{1/t}$ (again denoted by $u$) for
$\phi >\phi _c$
\begin{equation}
Au^{1+\alpha}-Au^{\alpha }-
u{(\phi-\phi _c)(1-\phi -\phi _c)\over \phi _c(1-\phi _c)}x_+^{1/s}-
x_+^{1/s} =0.
\label{xp}
\end{equation}
If the term linear in $u$ of Eqs.(\ref{xm}) and (\ref{xp}) could be
neglected, the
scaling functions $F_{\pm}$ would manifestly depend only on the respective
variables $x_{\pm}$. It is due to this term, that scaling
is invalidated to a certain degree by the solution for $\sm $ 
of Eq.(\ref{gem}).
The range and extent to which this is the case is discussed below. 
An interesting aspect of Eqs.(\ref{xm}) and (\ref{xp}) is exact 
scaling at $\phi =\phi _c$ and $\phi =1-\phi _c$. Whether or not exact 
scaling for $\phi =1-\phi _c$ is merely a coincidence can only 
be revealed by appropriate experiments. We stress
that all results obtained in this section are independent of whether the
conductances and hence the scaling functions $F_{\pm }$ are genuinely
complex or real.

Exact solutions of Eqs.(\ref{xm}) and (\ref{xp}) can be read off at the
limit points of the concentration. At
$\phi =0$, it is $u^{\alpha }=x_-^{1/s}$, i.e.$\,F_-\equiv x_-$ and
at $\phi =1$, it is $u=1$, i.e.$\,F_+\equiv 1$.
From these solutions, Eqs.(\ref{limm}) and (\ref{limp}) are obtained
from Eqs.(5) and (6), respectively. In fact, the respective solutions
are valid to high accuracy for $\phi>0$ and $\phi <1$ as long as 
$|x_{\pm }|\ll1$ or for $\phi $ very close to either $\phi _c$ or $1-\phi _c$. 
Correction terms are given below.

For the opposite limit of
the scaling parameters, i.e.$\,|x_{\pm}|\gg 1$, we obtain for the leading
term at $\phi \approx \phi _c$ the solution $Au^{1+\alpha }=x^{1/s}$ from both
equations, (\ref{xm}) and (\ref{xp}). This translates into 
\begin{equation}
\label{fphc}
F_{\pm}=(x_{\pm})^{{t\over s+t}}A^{-{st\over s+t}}
\end{equation}          which gives
\begin{equation}
\label{phc}
\sm ={\sc \over A^{{st\over s+t}}}
\bigl({\si \over \sc}\bigr)^{{t\over s+t}}
\quad {\rm at} \quad \phi=\phi _c.
\end{equation}
Note that Eq.(\ref{phc}) conveniently
lends itself for complex values of $\si $ and $\sc $; in particular, if
$\si $ is purely imaginary and $\sc $ real one obtains
\begin{eqnarray}
\label{imsm}
\Im \sm &=&
-{\sc \over A^{{st\over s+t}}} \bigl|{\omega \epsilon_0 \epsilon_I 
\over \sc }\bigr|^{{t\over s+t}} 
\sin ({\pi t\over 2(s+t)}),  \\
\label{resm}
\Re \sm &=&
+{\sc \over A^{{st\over s+t}}} \bigl|{\omega \epsilon_0 \epsilon_I 
\over \sc }\bigr|^{{t\over s+t}} 
\cos ({\pi t\over 2(s+t)}).  
\end{eqnarray}
From Eq.(\ref{fphc}) it follows that the slope of the real and imaginary
part of $\log (F_{\pm})$ is $t/(t+s)$ when plotted against $\log (x_{\pm})$
for $|x_{\pm}|\gg 1$. Equations(\ref{imsm}) and (\ref{resm}) show that, with 
$\si =-i\omega \epsilon_0 \epsilon_I$, the frequency dependence of both
real and imaginary $\sm $ is $\omega ^{{t\over s+t}}$. This dispersion law
is given in \cite{7.}-\cite{10.}. Experimental results validating this power
law are found in \cite{7.,12.},\cite{20.}-\cite{24.}. The loss angle
$\delta =\arctan [\pi t/(2(s+t))]$ implied by Eqs.(\ref{imsm},\ref{resm}) is
also given in \cite{8.}.

It is of physical interest to determine the correction terms of next order in
Eqs.(\ref{limm}) and (\ref{limp}). Note that Eq.(\ref{limm}) yields a purely
imaginary result for $\sm $ if $\si $ is imaginary. However, a loss term
should emerge for $\omega >0$ when $\phi >0$. This is
obtained by expanding the solution of Eq.(\ref{xm}) for $\phi \simgeq 0$.
One finds for the scaling function 
\begin{equation}
\label{realf}
F_-= x_--s{\phi \over \phi _c^2}x_-{x_-^{{1\over t}}\over
A x_-^{{1\over t}}+1}=
{\si \over \sc}\biggl(1+(s+t){\phi \over \phi _c}-
s{\phi \over \phi _c^2}{\bigl({\si \over \sc}\bigr)^{{1\over t}}\over
A\bigl({\si \over \sc}\bigr)^{{1\over t}}+1}\biggr)
\end{equation}
which can be used for real or complex $\si $ or $\sc $. Combining
Eq.(\ref{realf}) with Eq.(\ref{smless}) one obtains for small $\phi /\phi _c$,
as expected, both an enhanced dielectric loss term $\Re \si (1+s\phi /\phi _c)$
and a composite loss term. Taking specifically $\si $ purely imaginary and
$\sc $ real, the composite loss term reads explicitly up to terms linear in
$\phi / \phi _c $
\begin{equation} \label{losst}
\Re \sm =s\sc {\phi \over \phi _c^2}\,
{\bigl|{\si \over \sc }\bigr|^{{1+t\over t}}\sin ({\pi \over 2t})
\over A \bigl|{\si \over \sc }\bigr|^{{2\over t}}+
2A\bigl|{\si \over \sc }\bigr|^{{1\over t}}\cos({\pi\over 2t})+1}.
\end{equation}
An important consequence of Eq.(\ref{losst}) is the small frequency behaviour
of the loss term (recall $\si =-i\omega \epsilon_0 \epsilon_I$), 
which implies
\begin{equation}
\label{freq}
\Re \sm \sim \omega ^{{1+t\over t}},  \end{equation}
which differs from the $\omega ^2$-behaviour predicted by the expansions 
used for $F_-$ in \cite{8.}-\cite{10.}. We note that these expansions assume 
analytic behaviour for $\sm $ around $\omega =0$ which is in contrast to our
findings; also we obtain a loss term that vanishes for $\phi \to 0$, which is
not the case for the expressions in \cite{8.}-\cite{10.}. The following
section presents experimental results which appear to confirm the power
law expressed by Eq.(\ref{freq}).

By similar means we obtain the first
order correction term in the vicinity of $\phi \simleq 1$ which reads
\begin{equation}
\label{imf}
F_+=1+t{1-\phi \over \phi _c(1-\phi _c)}{x_+^{{1\over s}}\over
A+x_+^{{1\over s}}}=
1+t{1-\phi \over \phi _c(1-\phi _c)}
{\bigl({\si \over \sc }\bigr)^{{1\over s}}\over A+
\bigl({\si \over \sc }\bigr)^{{1\over s}}}.
\end{equation}
Note that this term implies not only
a correction to the real part of $\sm $ in Eq.(\ref{limp}) but also a
switching on of an imaginary part for complex $\si $. For purely imaginary 
$\si $ this is
\begin{equation}  \label{ref}
\Im \sm =-t{1-\phi \over \phi _c(1-\phi _c)} \sc
{\bigl|{\si \over \sc }\bigr|^{{1\over s}} \sin({\pi \over 2s})\over
A^2+2A\bigl|{\si \over \sc }\bigr|^{{1\over s}}\cos({\pi\over 2s})
+\bigl|{\si \over \sc }\bigr|^{{2\over s}} }
\end{equation}
which implies in this limit, for $\si =-i\omega \epsilon_0 \epsilon_I$, that 
$\Im \sm \sim \omega ^{1/s}$.

So far, we have concentrated on regions where scaling is obeyed by the
solution of Eq.(\ref{gem}) either exactly or to high accuracy, that is
for $0\le \phi < \phi _c$ and $\phi _c<\phi \le 1$, 
if $|x_{\pm}|\ll 1$, and for $\phi \approx \phi _c$,
if $|x_{\pm}|\gg 1$. There is an intermediate region $|x_{\pm}|\sim 1$,
where the linear term in $u$ of Eqs.(\ref{xm}) and (\ref{xp}) does invalidate
the sole dependence of $F_{\pm}$ on $x_{\pm}$ except for $\phi = \phi _c$
or $\phi = 1-\phi _c$. In fact, it can be shown that, as long as the
inequality
\begin{equation}
{\omega \over \omega_0} < {\phi _c \over 1-2\phi _c} \quad {\rm with} \quad
\omega_0 = \bigg|{\sc \over \epsilon_0\epsilon _I}\bigg| \label{ineq}
\end{equation}
is obeyed, the linear term of Eqs.(\ref{xm}) and (\ref{xp}) is immaterial 
and scaling prevails. As a consequence,
the leading behaviour of $F_{\pm }$, for $x_{\pm }\gg 1 $,
is governed by the power law $x_{\pm }^{t/(t+s)}$ only up to the frequency
which obeys the inequality (\ref{ineq}), for larger frequencies $F_{\pm }$
becomes a linear function of $x_{\pm }$.
Note, however, that the right hand side of Eq.(\ref{ineq})
depends on $\phi _c$ in such a way that scaling is expected to be invalidated
only for small values of $\phi _c$ and sufficiently large values
of $\omega $. In turn, for $\phi _c > 1/3$ the right hand side
of (\ref{ineq}) is larger than unity, and for $\phi _c \to 1/2$ no bound
on $\omega $ prevails. (Note that $\phi _c=1/3$ is the Bruggeman value for
spheres in three dimensions and $\phi _c=1/2$ for discs in two dimensions
\cite{16.,17.}.) As a consequence, there should be
no discernible deviations from scaling for $\phi _c > 1/3$.
To what extent these results are physically valid can
only be assessed by experiment. No experiments in this region appear to exist
and the situation is complicated by the fact that $\sc $ and $\si $ depend on
$\omega $ when $\omega $ becomes sufficiently large.

In Fig.1 we illustrate the behaviour of
the real and imaginary parts of $F_{\pm }(x_{\pm })$ for $s=1,\,t=2$ and
$\phi=\phi _c=0.16$, that is for the situation where scaling holds exactly.
As discussed above, deviations are marginal when $\phi $ is near to
$\phi _c$ and become noticeable only when the inequality (\ref{ineq}) is
appreciably invalidated. Note the equal slopes for large $x_{\pm }$ of
all four curves in accordance with Eqs.(\ref{imsm},\ref{resm}). Also,
since $t>s$, the imaginary parts are larger than the real parts; for $s=t$
all four curves would coincide asymptotically. For $s>t$ the real
parts would be larger than the imaginary parts but no such system has been
observed or predicted.

We interpret the dependence on $\si /\sc $ of the percolation loss term as
predicted by Eqs.(\ref{realf}-\ref{freq})
as follows: Consider a three dimensional
lattice, with ideal capacitors on nearly all the bonds but with a small
number of randomly distributed resistors embedded in the capacitor matrix,
either as isolated components or as small clusters, and choose
$C\omega < 1/R$. If a voltage is applied to two opposite faces of the
lattice, the displacement current in the lattice is determined almost
entirely by the capacitors. Therefore, as the
current in the resistors is evoked by a "fixed current source", which is
determined by the value of $1/(C\omega )$, the dissipation is proportional
to $I^2R$, i.e. the larger $R$ the more power is dissipated. This argument
can be extended to continuum systems, which qualitatively explains the
dependence on $\si /\sc $ in Eq.(\ref{realf}). The presence of the
exponent $t$, which determines the rate of increase of the conductivity
beyond $\phi _c$, would therefore appear to
have a role in the formation of the conducting clusters, which determine
the dissipation below $\phi _c$. However, below $\phi _c$,
the behaviour of the complex effective
conductivity is dominated by the imaginary component, which is primarily
determined by the interconnectivity  of the insulating medium, or more
specifically its ability of keeping the clusters of the conducting medium
disconnected below $\phi _c$, which in turn is characterised by the
exponent $s$.

We note that the treatment given in \cite{8.}-\cite{10.} has no $t-$ or
$s-$dependence for $x_-<1$ or $x_+<1$, respectively. However, our 
Eqs.(\ref{realf}) and (\ref{imf}) which are based upon Eqs.(\ref{xm})
and (\ref{xp})
indicate that the exponent $t$, which characterizes the formation of the
conducting backbone, continues to play a role for $\phi <\phi_c$; similarly,
the real dielectric constant is not independent of $s$ for $\phi >\phi_c$.

\section{Experimental Method}

The percolation system \cite{7.,11.,12.} which best exhibits the $(1+t)/t$
behaviour is a lightly poured powder of 55\%g Graphite 45\%
Boron Nitride  which is compressed, expelling air, in a
capacitive cell through the percolation threshold. As the percolation
threshold is at 0.124 (volume fraction of G), the insulator at and around
this point consists of  11.4\%BN and  88.6\%  dry air, which obviously has
a $\Re \epsilon$ close to one and a very low $\Im \epsilon $ term.
The dissipation in more
compacted systems, such as compressed pellets of G-BN \cite{7.} or
polyethylene-carbon black \cite{18.}, would appear to be dominated by the
dissipation in the dielectric component.
 
As the constructions of the cells and experimental procedures 
for the G-BN powder have been
adequately described for the dc and low frequency ac measurements in
\cite{11.,12.} as well as the dispersion measurements
$(\sm (\omega)$ for fixed $\phi $) in \cite{7.,12.}, they will
not be repeated here. Some experimental results are presented, which are
obtained by tumbling various conducting powders with larger wax coated
insulating grains. The conductor coated grains are then compressed into
discs \cite{19.}.

The new measurements presented
here are results of the real and imaginary parts of the conductivity
between $10^{-1}$ and $3.10^6$Hz, obtained using a newly acquired Novocontrol
Dielectric Spectrometer. This instrument is able to measure far smaller
loss components of the dielectric or insulating phase (equivalent to a
resistor of $10^{18}\Omega $ at $10^{-1}$Hz and $10^8\Omega $ at $10^5$Hz
in parallel with a
perfect capacitor) and had a better resolution of loss or phase angle (a
maximum of $\tan \delta $ of $> 10^3$ and a minimum of $< 10^{-3}$
can be measured) than
the instruments used in \cite{7.} or measurements of a similar nature
\cite{20.}--\cite{24.}.

\section{Results and Discussion}
The experimental results for $\Re \epsilon $ and $\Re \sm $ for the
55\%G-45\%BN powder as function of frequency between $10^{-1}$ and
$3.10^6$Hz are given in Figs.2 and 3.
The dispersion results for the three highest lying curves in Fig.3   are
conducting samples ($\phi >\phi _c$) 
with dispersion free conductivities  at low frequencies \cite{8.}-\cite{10.}. 
The dielectric constant of these three samples (upper curves in Fig.2)
show a strong dispersion,
which should go from $s/(s+t)=0.47/(0.47+4.8)\approx 0.09$ at $\phi _c$
to $1/s \approx 2$ near to $\phi =1$. The observed values range from
0.1 to 0.33. The dielectric constant for the conducting samples 
($\phi >\phi _c$) will be further discussed below.
       
The sample marked by plus signs in Fig.3, ($\phi -\phi _c =-0.0004\pm 0.001$) 
is actually metallic, as its conductivity  breaks away from a constant slope
at sufficiently low frequencies,
as predicted for metallic samples in this paper and \cite{8.}--\cite{10.}.
Note too that the larger exponent 0.15 for the higher frequencies
shown by this sample in Fig.2 also indicates that it is not an insulator
($x_- < 1$). When compared to the results in Fig.1 in \cite{7.},
this observation illustrates the necessity of making measurements at lower
frequencies than previously done in any experiments of this nature.

In Fig.2,
the exponent of $\Re\epsilon $ drops form -0.10 to -0.02 for the
next four samples. While, from the results given in
\cite{7.}, a dispersion exponent of -0.10 could indicate a sample still in
the $x_- > 1$ region, the fact that the sum of the  absolute values of  the
exponents for  the dispersion in $\Re \epsilon$  and $\Re \sm $ are greater
than one, precludes this possibility \cite{7.}--\cite{10.}. This, combined
with the ever decreasing magnitude of the $\Re\epsilon $ exponents, going down
to 0.02 from the lowest lying curve, allows us to conclude that these four
samples lie in the dielectric or $x_- <1$ region. Note that the lowest
measured values for  $s/(s+t)$ recorded in the literature
\cite{7.},\cite{19.}--\cite{23.} are 0.07 for 50\%G-50\%BN and 0.10 for
55\%G-45\%BN \cite{7.}.

Based on the arguments given above, the four lowest lying plots of the 
conductivity against
frequency in Fig.3 are the conductivities for samples in the insulating or
dielectric state ($x_- < 1$).  There is a very slight upward curvature of
the results between 10 and $3.10^6$Hz, but the mean slopes in this region are
1.06 and 1.09. The slope decreases below
3Hz. According to Eq.(\ref{realf}) the conductivity is made up of both a
dielectric and percolation loss term, the relative contributions of which
must now be examined.

In order to evaluate Eqs.(\ref{realf},\ref{losst}),(\ref{imf})
and (\ref{ref}), an expression for $\si $ is
required. Unfortunately the loss term in the Boron Nitride-Air system is
too small to measure directly on the dielectric spectrometer, which also meant
that the percolation parameters for this system could not be measured
directly. Therefore, the complex conductivity of  Boron Nitride-Air mixture
had to be calculated, using effective media theories, from measurements made
on a compressed disc with a porosity of 0.19. When plotted against the
frequency, in the range $10^{-1}$ to $3.10^6$Hz, $\Re\epsilon $ for this disc
was found to be virtually constant, and that the $\Re\sm $ term could be 
fitted to $\Re\sm =\sigma _{{\rm dc}}+D \omega ^{0.9}$ \cite{25.} with
$\sigma _{{\rm dc}}$ being sensitive to how dry the BN in the disc was (and
therefore also how dry the BN in the G-BN powder was). By pumping on the
discs long enough the dc component became unmeasurable
(i.e. $<10^{-18}(\Omega {\rm m})^{-1}$
at $10^{-1}$Hz). $D$ was independent of $\sigma _{{\rm dc}}$ and had a value
of  $9.0\times 10^{-16}$. Unfortunately, the powders could not be pumped 
to lower values of $\sigma _{{\rm dc}}$ as this caused them to collapse.

Although the system is anisotropic \cite{7.}, if measurements are made in
the axial (compression) and radial directions, it is probably still valid to
use the Hashim-Strickman (H-S) upper and lower bounds \cite{26.} for
measurements made in the radial direction only. Therefore, $\Re\sm $ and
$\Im\sm $ for "bulk" BN where determined using the upper bound, as the BN
grains are obviously in contact at low volume fractions, which makes the
system closer to one where the BN surrounds the air \cite{27.}. The bulk
parameters are $\Re\epsilon  =4.1$ and $D=1.2\times 10^{-15}$. As the
critical
volume fraction for the 55\%G-45\%BN system is 0.124, the upper limit for
dilute systems was taken to be $\phi = 0.02$. At this volume fraction of G
the volume fraction of BN was 0.0164 and of air 0.9636.  Therefore the
parameters determining $\sm $ for a system with 0.017 volume fraction of BN
and  0.983 for air were calculated, again using the formula for the H-S
upper bound, which gave a $\Re\epsilon =1.03$   and a $D= 1.41.10^{-17}$
with $\sigma _{{\rm dc}}$ selected to fit the experimental results. The
calculated $\Re\sm $  at $\phi =0.02$ is then used to evaluate the
enhanced dielectric loss term, the percolation contribution and the
combination of these, using Eqs.(\ref{smless}) and (\ref{realf}). These are
all shown in Fig.3. The computed value of $\Re \sm $ for the Air-BN
insulator is not shown but lies below the enhanced curve
by a factor of $s\phi /\phi _c$. The other values $s= 0.47,\, t=4.8$   and
$\sc = 3126(\Omega {\rm m})^{-1}$, used to calculate these curves,
have all been determined by dc percolation
measurements \cite{11.,12.} and $\sigma _{{\rm dc}}$ was chosen to match the
curvature at low frequencies. The mean slopes of  1.05 for the squares
and 1.09 for the crosses are lower than the value of (1+4.8)/4.8 = 1.21,
as is predicted by Eq.(\ref{freq}). The discrepancy is explained by the
comparatively large values of $\phi $ in the experimental results;
in fact numerical solutions of Eq.(\ref{expl}) show that the
$\omega ^{1/(1+t)}$ behaviour is flattening out for increasing $\phi $ as
it has to attain the $\omega ^{t/(s+t)}$ behaviour for $\phi \to \phi _c$.

The only data that claims an $\omega ^2$ dependence for 
$\sm ,\, \phi <\phi _c$, is that of Benguigui \cite{21.}.  
However there are problems with his data in a mixture of iron 
balls and glass beads.  
His sample contains only $10^5$ particles (about $10^8$ in the present 
experiments not even 
counting the air volume) and only 30 particles between the 
capacitor plates (nearly 
300 plus air in the present experiments).  Although Benguigui 
mentions the dc conductivity of the glass, it would appear that neither 
this nor the dielectric loss term in the glass is taken into account. The
authors are also at a loss to explain how the very low frequency dispersion 
observed for $\Re \epsilon (\omega )$ in an insulating 20.0\% iron balls
sample (Fig.2 in \cite{21.}) can give a $\Re \epsilon (\omega )\omega 
\sim \Im \sm (\omega )$
which varies as $\omega ^2$ when plotted against the 
frequency (Fig.8 in \cite{21.}).  Therefore we do not regard these 
experiments as definitive.

In Fig.2 the dielectric results for the conducting samples 
are terminated at 1 Khz as for lower frequencies $\tan \delta $ 
exceeds $10^3$ and the dielectric spectrometer gives spurious 
results.  However, the results at 1 Khz (a frequency commonly used in 
low frequency  experiments to measure $\Re \epsilon $) 
clearly show that $\Re \epsilon $ continues to increase with $\phi $ below and 
above $\phi _c$.  This is in sharp disagreement with the predictions given in 
\cite{8.}-\cite{10.}, where it is claimed that $\Re \epsilon $ 
should decrease as $(\phi-\phi _c)^{-s}$ for $\phi>\phi _c$.

The smooth behaviour of $\Re \epsilon $ as a function of $\phi $ 
passing and extending beyond $\phi _c$ has 
also recently been observed in carbon black-polyethylene composites \cite{18.} 
and a 
number of systems where various fine conducting powders are 
impregnated onto the 
surface of almost spherical insulator grains before the coated grains 
are compressed 
into a three dimensional continuum \cite{19.}.  Therefore there is now strong experimental 
evidence that the second order term for $\phi >\phi _c$ 
given in \cite{8.}-\cite{10.} is in disagreement with 
experimental evidence.  Their second order percolation term also fails to 
vanish for $\phi \to 1$ in contrast to our result in Eq.(\ref{imf}).

However, Eqs.(\ref{gem}) and (\ref{expl}) show a $\Re \epsilon $ 
that continues to increase above $\phi _c$. 
Unfortunately, the agreement with the experimental results 
is qualitative, at best, if the 
parameters obtained from dc experiments are substituted into Eq.(\ref{expl}), 
as is shown in Figs.4 and 5.

Figure 4 shows the experimental $\Re \epsilon $  
results at 1 Khz and 1 Mhz, plotted against $\phi $, for 
the G-BN powder. All theoretical curves, calculated from Eq. (\ref{expl}) 
are for $t =4.8 $ \cite{11.}. 
The solid curves are for $\sc =3.1\cdot 10^3 (\Omega {\rm m})^{-1}$ 
\cite{7.,11.}, $\Re \epsilon=1.18 $ (calculated $\phi _c$ value), 
$\phi _c = 0.124$ and $s=0.96$. The upper solid curve is for 1 Khz and the 
lower one for 1 Mhz. The dotted 
curves use the same parameters except that $s= 0.6$. As 
the experimental results all lay 
above the theoretical curves, we display the dashed curve where 
$\sc $ has been changed to $3.1\cdot 10^5 (\Omega {\rm m})^{-1}$. 
From this figure and the 
behaviour of the theoretical plots it is apparent that  
the experimental results can be better fitted, if
some or all of the above parameters are varied to
get the best fit. In this case it 
would not be necessary to change $\phi _c$ from its dc value of 
$0.124 \pm 0.001$. As the dc
conductivity changes by nine orders of magnitude between 0.120 and 0.130, one 
cannot argue that the incorrect $\phi _c$ has been identified.

Of the six powders that were coated onto the wax coated insulating grains, as 
previously described, only the Nickel powder did not show an 
increasing $\Re \epsilon $ above 
the dc value for $\phi _c$, this was because the 
conductivity at and above $\phi _c$ was too high to 
measure $\Re \epsilon $ \cite{19.}. This combined with the 
results for carbon black-polyethylene \cite{18.}, 
both as a function of $\phi $, and $\phi _c$ (as a function of temperature)
for fixed $\phi $, show that this could well be the usual 
behaviour for continuum systems. The reason that this has 
not been previously reported is 
probably due to the limitations of $\tan \delta $ in the 
instruments previously used, and the fact 
that for $t\approx 2$  and large $\sc /(\omega \epsilon_0 \epsilon)$ ratios 
the solution of Eq.(\ref{expl}) is sharply 
peaked {\it just above} $\phi _c$.  To observe this difference accurate 
measurements of both the dc and the apparent ac value of $\phi _c$ 
(delayed peak in $\Re \epsilon $) would have had to be made. This has 
never been done except for the G-BN system, the coated grain system \cite{19.} 
and to some extent the carbon black - polyethylene system \cite{18.,28.}. 
In all the coated grain 
systems sharp changes in the dc curves of $\sm $ give unambiguous $\phi _c$ 
values \cite{19.}. With a poor 
conductor (magnetite) $\Re \epsilon $ increases sharply near $\phi _c$ and 
then smoothly up to $3\phi _c$ \cite{19.}.

Figure 5 shows the experimental results for $\Re \epsilon $
for a system of insulating grains coated 
with NbC \cite{19.}. All theoretical curves, calculated from Eq.(\ref{expl}) 
are for $t=4.8$. The solid curves are for 
$\sc = 7000 (\Omega {\rm m})^{-1}, \Re \epsilon =7.45, \phi _c = 0.065$ and 
$s = 0.8$. The upper 
solid curve is for 1 Khz  and the lower one 1 Mhz. The lower 
dotted curves are for the same parameters but $s = 0.4$, which is the measured
dc value. The low value for $\sc $ could 
be due to the resistance of the system being largely determined 
by contacts between the 
extremely hard and angular NbC grains. All the above parameters, 
except for $s=0.8$, are close to those obtained from the dc conductivity
fit using Eq.(\ref{gem}). Except for the fact that the drop in $\epsilon $ 
for larger values of $\phi $ is not observed the results could be called 
qualitatively correct.

The curves selected do not show that the theoretical 
curves widen considerably as $t$ is 
increased, i.e. the conductivity exponent plays a large 
role in determining $\Re \epsilon $. An explanation may
be that as the conductivity of a percolating system
above $\phi _c$ increases more slowly with $\phi $
for high $t$ values, this means that a smaller fraction 
of the conducting component is on the 
backbone of a system with a higher $t$ value than a system with a lower $t$. 
This "off the backbone" conducting material then creates nearly 
conducting links between different sections of the backbone, which are 
broken by the insulating component. It would  then be the effect of these
inter {\it dead end} capacitances that continues to increase $\Re \epsilon $
above $\phi _c$, until these capacitances are shorted out by an ever more
conductive backbone.

\section{Summary and Conclusions}

In this paper we have shown analytically that Eq.(\ref{gem}) in 
conjunction with Eqs.(\ref{smless}) and (\ref{smplus}) gives complex 
scaling functions for continuum percolation type 
systems with the following results:

i)  $\Re F_+(x_+)$ has zero slope for $x_+ < 1$ (first order term),  
and  a slope of  $t/(s+t)$ for $x_+ > 1$,

ii)  $\Im F_-(x_-)$  has slope unity for $x_- < 1$ (first order term), 
and a slope of t/(s+t) for $x_- > 1$.

iii)  A slope of  $t/(s+t)$ for $\Im F_{\pm }$ and $\Re F_{\pm }$ for $x_+$ 
and $x_- > 1$.

Note that for $x_+$ and $x_- > 1$ one cannot clearly distinguish 
between first and second 
order terms.   All of these limiting slopes agree with those 
given in \cite{8.} -\cite{10.}. However 
the $F_{\pm }$ in this paper are analytic functions with no 
unspecified constants. The 
functions $\Re F_+$ and $\Im F_-$ have been shown to continuously fit 
the first order dispersion data for the G-BN systems over the whole range
of values for $x_{\pm }$ \cite{7.}, and $F_{\pm }$ the dc conductivity
results for $Al_xGe_{1-x}$ \cite{14.}.

The slopes for $\Im F_+$ and $\Re F_-$ for $x_{\pm } < 1$ (second 
order terms) differ from those 
given in \cite{8.} - \cite{10.}, which are based on the plausible 
assumptions made in \cite{1.} -\cite{6.}. 
However, no definitive experimental verification  
of these two terms seems to exist 
and the new experimental work given here strongly 
favours the second order term for $\Re F_-$ 
given by Eqs.(\ref{gem}) and (\ref{smless}). The experimentally 
observed increases in $\Re \epsilon $  with $\phi $ 
above $\phi _c$, some of which are given in Figs.4 and 5, is completely 
incompatible with the 
percolation equations, as given in \cite{8.} - \cite{10.}, 
but is qualitatively in agreement with Eq.(\ref{gem}) if the 
separately measured dc parameters are used.  Better agreement between 
Eq.(\ref{gem}) and experiment could be obtained if best fit 
parameters were used.

From the above evidence Eq.(\ref{gem}) may well be a 
better description of the ac and dc 
conductivity (dielectric constant) of percolative type systems 
than the standard 
percolation equations given in \cite{8.} - \cite{10.}. However, as 
Eq.(\ref{gem}) is a 
phenomenological equation its validity may need to be further tested 
by experiment as must the standard percolation equations.

\bigskip 

\noindent
{\bf Acknowledgements} \par \noindent
One of us (DSM) would like to thank Professor B. I. Schlovskii for pointing
out the necessity of taking the loss component of the insulating component
into account and Professor A.K. Jonscher for a useful correspondence on low
loss dielectrics.

\vskip 1cm   \noindent

{\bf Figure Captions.}   \\

{\bf Fig. 1}   \noindent

Plots of  $F_+$ and $F_-$ against $x_+ (\omega /\omega _c^+)$ 
and $x_- (\omega /\omega _c^-)$, respectively. 
The parameters used are $\phi \approx \phi _c = 0.16, s=1$ and $t=2$. 
As $\phi \approx \phi _c$ the values used for 
$\sc $  must be accordingly large to yield a finite $\omega _c^{\pm }$. The 
upper solid curve is $\Re F_+$ and the lower one $\Re F_-$ 
(the second order dielectric loss term). 
The dashed line is $\Im F_-$ (the first order term below $\phi _c$), and 
the dotted line is $\Im F_+$. Note how this term rises above 
$\Im F_-$ in the region where $x_{\pm }$ is between 1 and 100. 

{\bf Fig. 2 } \noindent 

A plot of the real part of the dielectric constant against frequency 
for a 55\%-45\% G-BN powder on a log log scale for various values 
of $\phi $ ($\phi $ = 0.1309 (open circles), 0.1290 (triangles),  
0.1272 (open squares), 0.1236 (plus), 0.1219 (crosses), 0.1203 (dots), 
0.1187 (asteriks), 0.1171 (solid squares)). These are 
relative volume fractions, as the absolute error is about $\pm 0.001$.

{\bf Fig. 3 } \noindent 

A plot of the real part of the conductivity against frequency for 
a 55\%-45\% G-BN powder on a log log scale for various values of 
$\phi $ ($\phi $ = 0.1309 (open circles), 0.1290 (triangles), 
0.1272 (open squares), 0.1236 (plus),0.1219 (crosses), 0.1203 (dots), 
0.1187 (asteriks), 0.1171 (solid squares)). These are 
relative volume fractions as the absolute error is about $\pm 0.001$.

{\bf Fig. 4 } \noindent 

Experimental values of $\Re \epsilon $ for G-BN, plotted against 
$\phi $, at 1Khz (dots) and 1Mhz (circles). The nature and parameters for 
the theoretical curves are given in the text.

{\bf Fig. 5 } \noindent 

Experimental values of $\Re \epsilon $  for NbC coated grains, plotted 
against $\phi $, at 1Khz (dots) and 1Mhz (circles). The nature and 
parameters for the theoretical curves are given in the text.

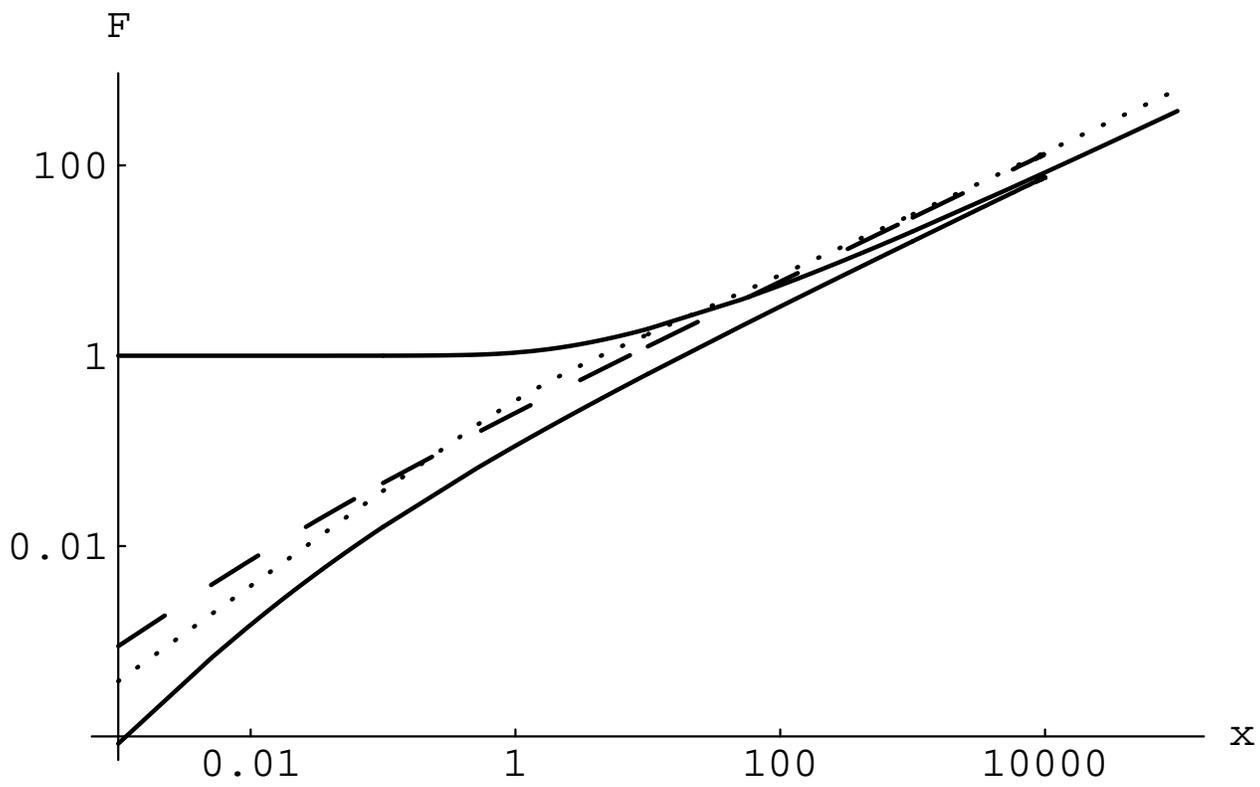
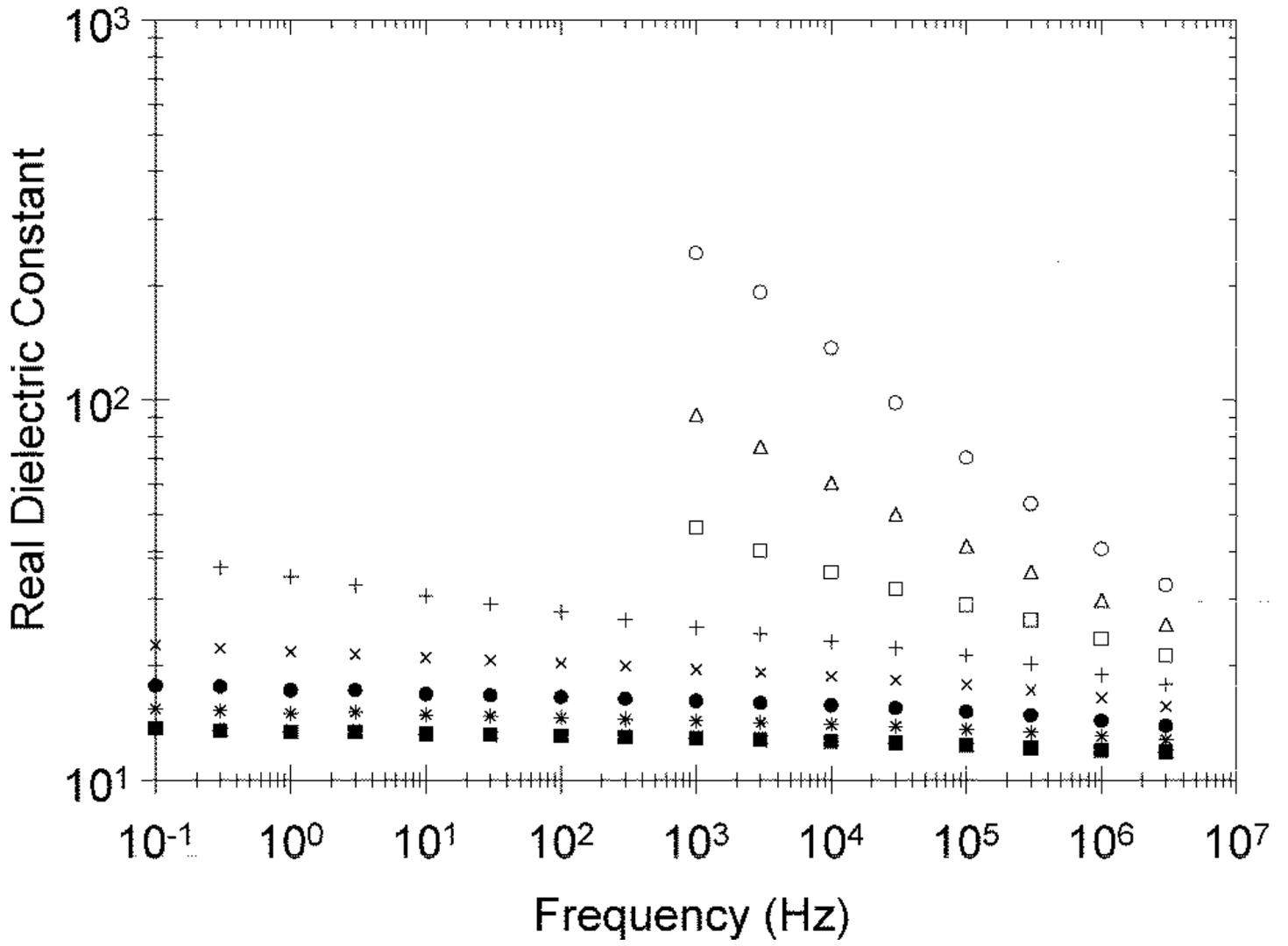
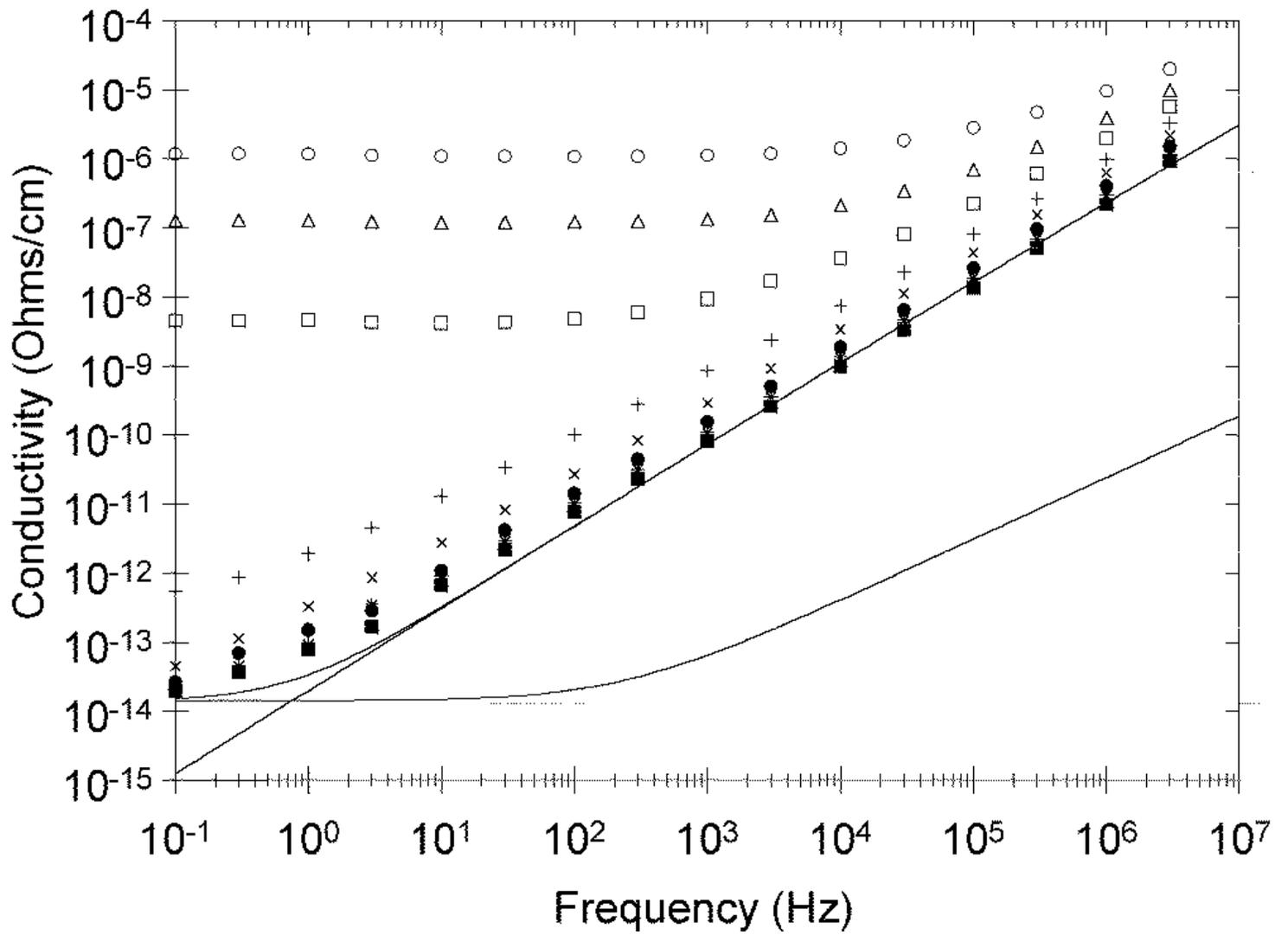
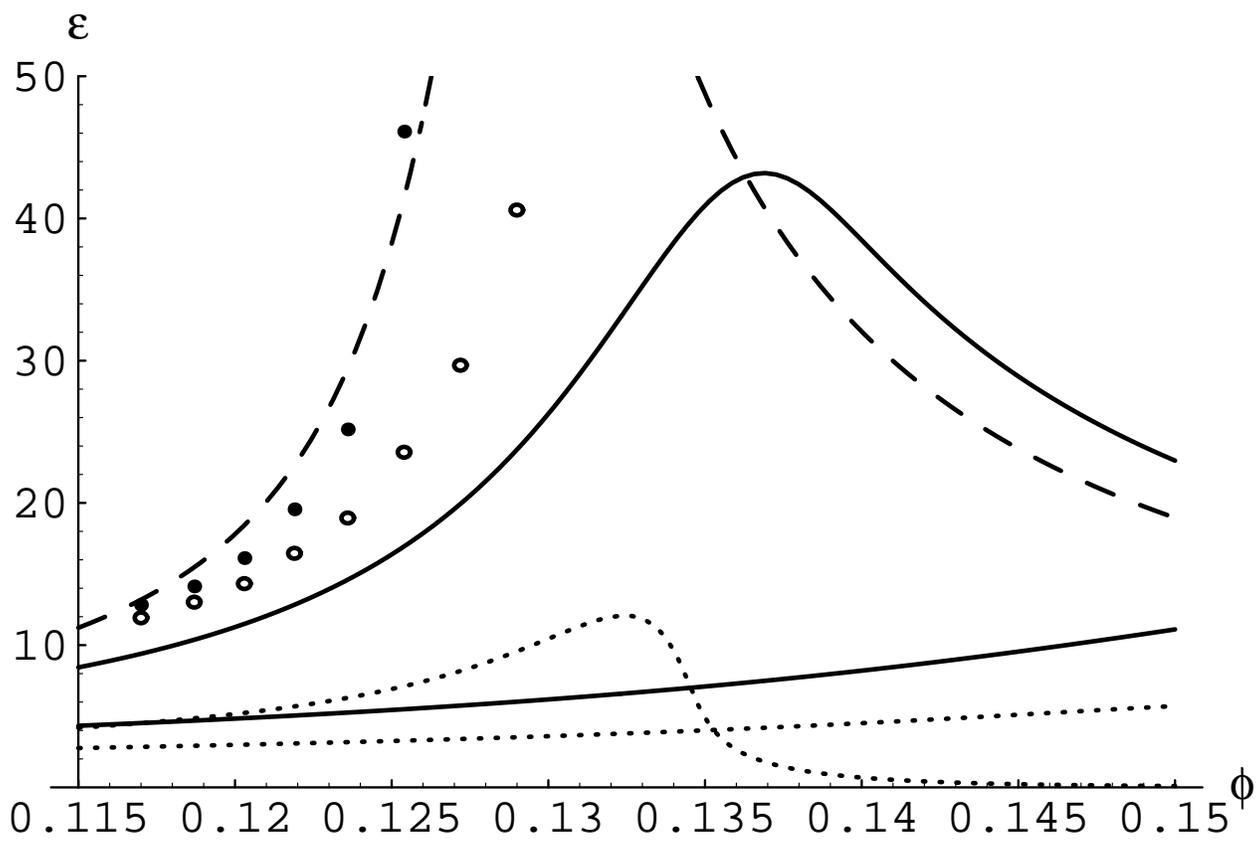
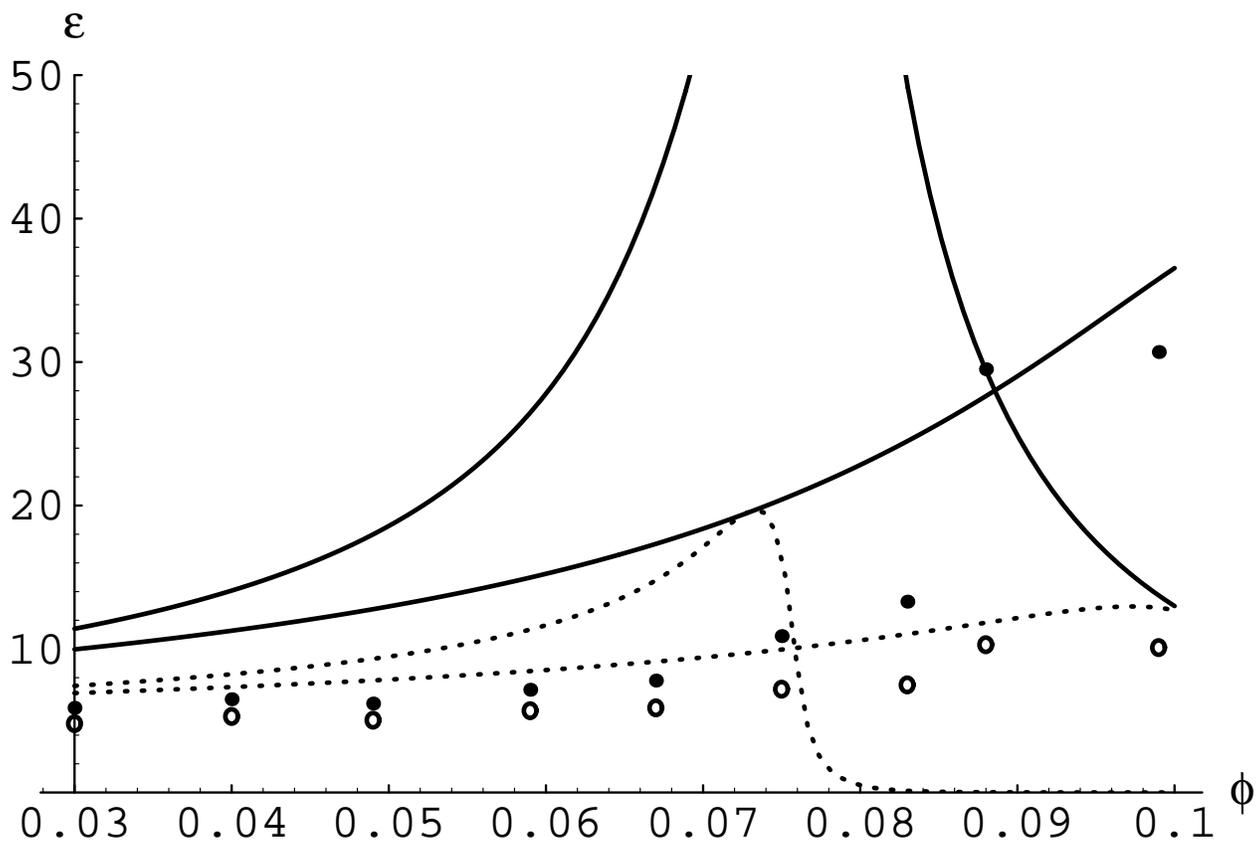
\end{document}